\documentclass[preprint,12pt]{elsarticle}

\usepackage{amssymb}
\usepackage{amsmath}
\usepackage{lipsum}
\usepackage{xcolor}
\journal{Optics Communications}

\begin{document}

\begin{frontmatter}



\title{Phase control of transmission and reflection in a sample of duplicated
two-level systems driven by a stationary control field}


\author[label1,label2]{F. A. Hashmi}
\author[label2]{E. Brion}
\author[label2]{M. A. Bouchene}

\affiliation[label1]{organization={Department of Physics, Syed Babar Ali School of Science and Engineering, Lahore University of Management Sciences (LUMS)},
            %addressline={},
            city={Lahore},
            %postcode={},
            %state={},
            country={Pakistan}}

\affiliation[label2]{organization={Laboratoire Collisions Agrégats Réactivité, FeRMI, Université
de Toulouse and CNRS UMR 5589},
            %addressline={},
            city={Toulouse},
            %postcode={},
            %state={},
            country={France}}

\begin{abstract}
In this article, we %investigate 
study the optical response of a duplicated two-level atomic medium %submitted 
subjected  to a stationary control field and a weak co-propagating probe field, orthogonally polarized to each other. We show that both the  reflected and transmitted components of the probe %may 
can  be absorbed or  amplified. %Moreover
Furthermore, for moderate optical depths, the  reflection and transmission factors are controlled by the relative phase between the control and probe fields, %which makes 
making  the configuration we present here promising for the development of optical devices. %Moreover, we 
We also  determine the exact conditions %where 
under which  reflection and transmission factors can be controlled by the relative phase between the  control and probe fields.
\end{abstract}



\begin{keyword}
%% keywords here, in the form: keyword \sep keyword, up to a maximum of 6 keywords
coherent control \sep four-wave mixing \sep nonlinear optics \sep duplicated two-level system



\end{keyword}

\end{frontmatter}

\section{Introduction}
The control and manipulation of light pulses has long been a key challenge
with many applications, %in, 
e.g.,  in  telecommunications and more recently in
quantum technologies. To achieve this goal one %may 
can try %and 
to  shape the
optical response of the propagation medium -- absorption and /or
dispersion -- %through 
by  combining %non-linear 
nonlinear  effects and quantum interferences.
%Electromagnetic 
Electromagnetically  Induced Transparency (EIT) \cite{BIH91}, %refraction
refractive  index enhancement \cite{Scu91}, slow, stored and fast light \cite{Hau99}
are examples of applications of such an approach. %Meta-materials like negative-index 
Metamaterials such as negative index  samples \cite{Pen00} or %photonics 
photonic  crystals \cite{JJW08}
are based on different strategies. In the latter, the propagation
of light is modified by the existence of photonic band gaps due to
periodic spatial variations of the optical index, in the same way
%as 
that  periodic atomic lattices affect the conductivity of electrons in
semiconductors. Such crystals are usually obtained by periodically  stacking dielectric
slabs %of 
with different indices. %periodically. 
The optical %features 
properties of the
resulting  structures %thus obtained 
are determined once and for all. Another
more versatile way %to obtain 
of obtaining  a photonic crystal is %provided by 
by means of  Electromagnetically
Induced Gratings (EIG). In this case, the periodic modulation of the
%medium 
optical response of the medium results from the application of a standing-wave
driving field, %which gives 
giving  rise to new original phenomena. For %instance
example,
the standing-wave configuration was recognized years ago %to be 
 as a  very
promising  way  to achieve spatial localization \cite{HSZ97}. Moreover,
 by   exciting a lambda system in EIT configuration with a stationary control
field one obtains a periodic modulation of the atomic absorption with
sharp peaks which allows %for the creation 
 the generation   of stationary light pulses %of light 
\cite{AL02} as well as the optical control of photonic band
gaps \cite{SH05}. Many groups also use this configuration to control
the group velocity \cite{JZW16}, to induce Raman gratings \cite{AM14},
phase gratings \cite{Ara10}, or to  implement phase gates and optical
switching \cite{SRZ11}. Multi-wave mixing can also be used to generate optical gratings to induce controllable dipole-mode solitons in three-level atomic systems \cite{ZWN11}  and edge-solitons in photonic graphene \cite{ZWZ20}.

In the present work, we focus on an atomic medium with the so-called
duplicated two-level system (DTLS) configuration,  which is  often used to study
electron spin coherence effects \cite{LWK03}. When excited by a
strong control and a weak probe beams, polarized orthogonally to each
other, DTLS exhibits efficient quantum interferences between  the   absorption
and stimulated emission paths. We pointed out the great potentialities
of DTLS media for controlling the propagation of the probe field in
previous works. We experimentally demonstrated the coherent control
of the medium gain in the femtosecond regime \cite{DB07}. We also
theoretically predicted the existence of Zeeman-coherent-oscillation-assisted
%slowing 
deceleration  of the probe in a non-collinear configuration \cite{HB08a}
and phase control of the effective susceptibility in a collinear configuration
\cite{HB08b}. Other authors proposed to take advantage of DTLS media
to control optical bistability \cite{WLZ10}, group velocity \cite{AS14},
and  to   implement quantum memory or optical switching \cite{AC10}. It
was also suggested to use orthogonal spatial configuration for the
control field to generate spatial diffraction within the multiphoton
resonance condition \cite{YPS13} or beyond \cite{VS18}. The phase
control of spatial interference of resonance fluorescence \cite{JN11},
transmission and reflection factors in a dielectric slab \cite{Sma23},
and the possibility of spatial localization of atoms \cite{ZYX16}
are among  the  promising related results. 

In the present article, we consider a DTLS medium %submitted 
 subjected   to collinear
control and probe fields. The control field %, however, 
 is now a standing-wave
%which induces 
 inducing  an EIG in Zeeman coherences. The overlap between the
control and probe beams ensures efficient interaction and energy exchange
between the field components along the sample. %By
 In  contrast  to %with
transverse gratings which only diffract a weak part of the probe field,
the present DTLS medium leads to significant reflection and
transmission coefficients which can  even  be %even larger 
 greater  than one. Another significant advantage 
%of the present configuration 
over 
%those using 
non-degenerate $\Lambda$ \cite{ZZW11} or other multilevel %systems 
\cite{UWG14, DD24} configurations, is that the implementation of the DTLS scheme %needs 
requires only one laser source %for its experimental implementation 
and phase matching is automatically satisfied for both reflected and transmitted beams. However, propagation effects affect both the control and the probe fields in DTLS. The aim %scope 
of the present study is to determine the exact quantitative conditions %where 
under which %a 
proper phase-control of the reflection and %the
transmission factors can be achieved in this system.
 In addition, %Moreover, 
we study the influence
of  the  interaction parameters -- control field intensity, detuning, optical
depth and phase shift between %the 
control and %the 
probe -- on  the intensity and phase of the  reflected
and transmitted probe light. %intensity and phase.  
\section{Theoretical model}

\subsection{The system }

\begin{figure}
\begin{centering}
\includegraphics[width=\textwidth]{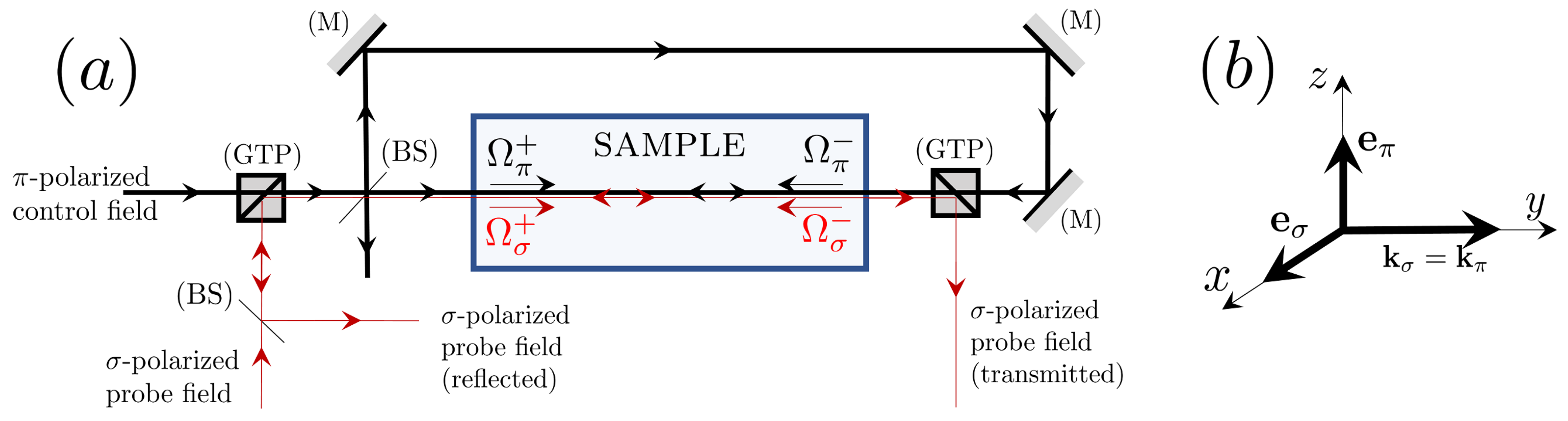}
\par\end{centering}
\caption{(a) Experimentally feasible configuration. The control field is split
into two parts that give the forward and backward components (Rabi
envelopes $\Omega_{\pi}^{\pm}$, respectively). The probe field is
sent  to the input %at the entrance 
of the sample in the forward direction (Rabi
envelope $\Omega_{\sigma}^{+}$) and the backward component (Rabi
envelope $\Omega_{\sigma}^{-}$) is generated from the  output   %end part 
of the sample. Here, (M) stands for mirror, (BS) for beam splitter, (GTP) for Glan-Taylor Polarizer.  (b) Field configurations. \label{Fig1}}
\end{figure}

We consider the experimentally feasible configuration %depicted 
 shown  in Fig. \ref{Fig1}.
An atomic sample of length $L$ is %submitted 
 subjected  to two fields of %same
 equal  frequency $\omega$, propagating along the $y$ axis with orthogonal linear
polarizations. The first -- so-called control -- field is $\pi$-polarized, i.e. its polarization is along the arbitrarily chosen quantization $z$-axis. It is
%split 
 divided  into two parts: one part %penetrates 
 enters  the sample from the
left side (%henceforth designated as 
 hereafter called   the  input), and propagates
in the direction of increasing $y$\textquoteright s, while the other
part is seeded back into the sample from the other side (%henceforth designated as 
 hereafter called  the output), and propagates in the direction of decreasing
$y$\textquoteright s. The total control electric field  is expressed as
\begin{equation}
{\bf E}_{\pi}\left(y,t\right)={\bf e}_{z}\left[\varepsilon_{\pi}^{+}\left(y\right)e^{\text{i}ky}+\varepsilon_{\pi}^{-}\left(y\right)e^{-\text{i}ky}\right]e^{-\text{i}\omega t}+\text{c.c.}\label{Control}
\end{equation}
For %sake of 
simplicity, we assume that the amplitude $\varepsilon_{\pi}^{+}\left(y=0\right)$
is real. Note that the $\pm$ subscripts in $\varepsilon_{\pi}^{\pm}$ refer to the %sense 
 direction  of propagation along the $y$-axis, either forward $(+)$ or backward $(-)$. The second -- %so-
called  the  probe -- field is $\sigma$-polarized, i.e. here  it is linearly-polarized along the $x$-axis and its
%propagates in the direction of increasing $y$\textquoteright s and
%penetrates the sample through the entrance. Its 
electric field is given by
\begin{equation}
{\bf E}_{\sigma}\left(y,t\right)={\bf e}_{x}\left[\varepsilon_{\sigma}^{+}\left(y\right)e^{\text{i}ky}+\varepsilon_{\sigma}^{-}\left(y\right)e^{-\text{i}ky}\right]e^{-\text{i}\varphi}e^{-\text{i}\omega t}+\text{c.c.}\label{Probe}
\end{equation}
In this expression, $\varphi$ is defined as the phase shift between
the probe and control fields. Note that even if the %injected 
probe field is originally injected through the  input  %entrance 
and
propagates in the forward direction, a backward component $\varepsilon_{\sigma}^{-}\left(y\right)$
 is built %builds 
up because the atomic sample radiates in both directions.  Furthermore, %Moreover,
the boundary conditions for the probe at the  output and input  %exit and entrance 
write
$\varepsilon_{\sigma}^{-}\left(y=L\right)=0$ and $\varepsilon_{\sigma}^{+}\left(y=0\right)=E_{0},$
where $E_{0}$ is real. Note that 
%Again, we want to emphasize here that 
the $\pm$ subscripts in $\varepsilon_{\sigma}^{\pm}$ 
%refer to the sense of propagation along the $y$-axis of the linearly-polarized probe field, and 
must not be confused with the sense of %a 
circular polarization.

\begin{figure}
\begin{centering}
\includegraphics[width=0.7 \textwidth]{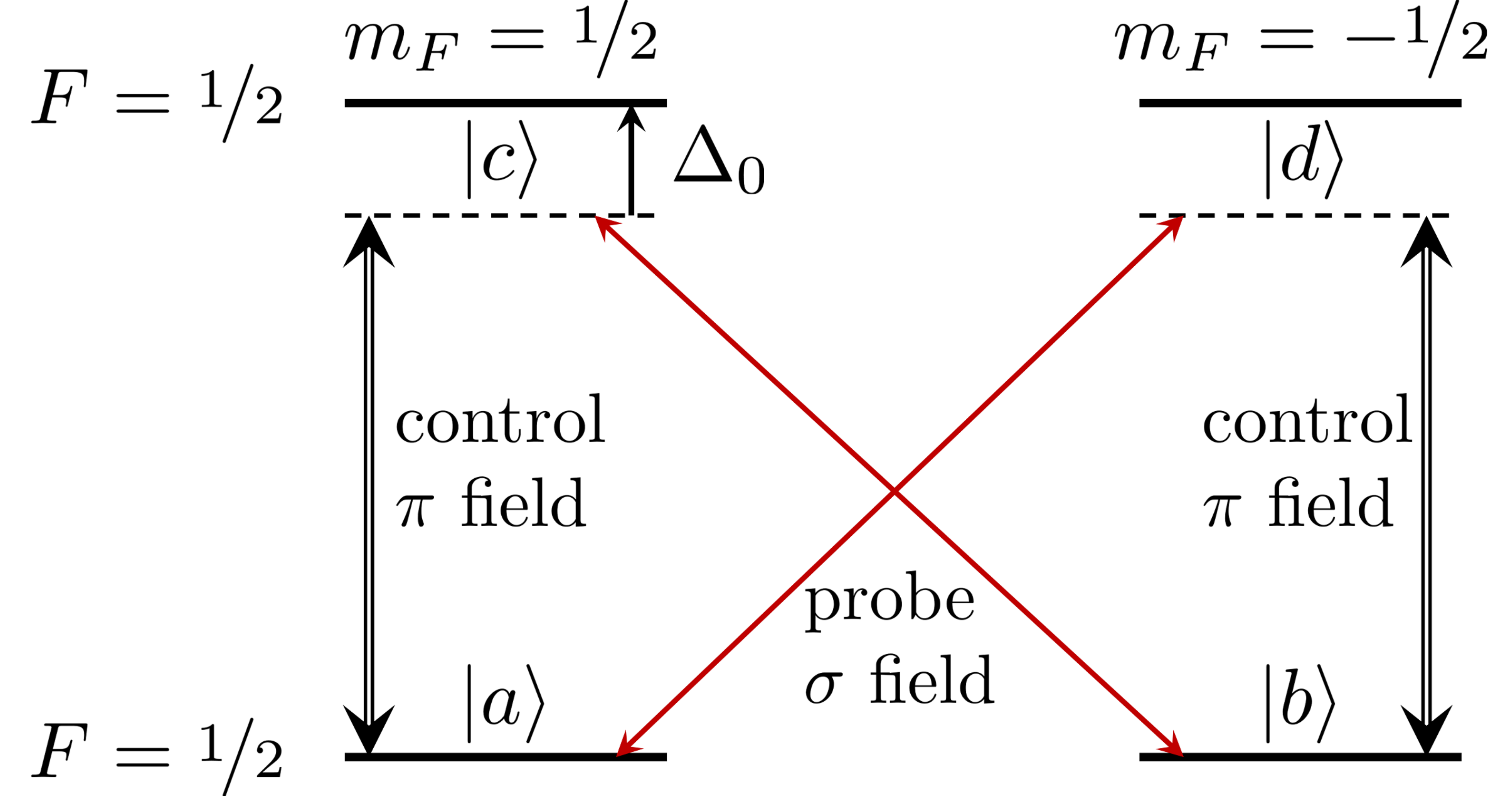}
\par\end{centering}
\caption{The duplicated two-level system (DTLS).
\label{Fig.2}}
\end{figure}

We assume that the fields  are  %essentially couple
 coupled to a $F=\frac{1}{2}\rightarrow F=\frac{1}{2}$
atomic line of the medium (e.g. $^{2}S_{\frac{1}{2}}F=\frac{1}{2}\rightarrow{}^{2}P_{\frac{1}{2}}F=\frac{1}{2}$
transition of $^{6}$Li at $671$ nm)  which is  described by a duplicated two-level
system (Fig. \ref{Fig.2}). The $\pi$-polarized control field ${\bf E}_{\pi}$
(resp.  the  $\sigma$-polarized probe field ${\bf E}_{\sigma}$) couples
to the $\Delta m_{F}=0$ (resp. $\Delta m_{F}=\pm1$) paths. Note that light hole transitions in quantum wells may also be considered \cite{LWK03}. In the
following, we determine the reflection and transmission coefficients
of the medium for a weak probe in the presence of a much stronger
control field. 

\subsection{Atomic polarization}

In the rotating frame defined by the unitary transformation $U\left(t\right)=e^{-\text{i}\omega t\left( \left| c \right\rangle \left\langle c \right| +\left| d \right\rangle \left\langle d \right| \right)}$
%In the interaction picture, 
the density matrix $\rho$ of an atom in the medium obeys the master equation $\text{i}\hbar\partial_{t}\rho=\left[H,\rho\right]+\text{relaxations}$
where 
%$H$ is 
the Hamiltonian takes the following form 
\[
H=\hbar\left(\begin{array}{cccc}
0 & 0 & -\Omega_{\pi}^{*} & -\Omega_{\sigma}^{*}e^{\text{i}\varphi}\\
0 & 0 & -\Omega_{\sigma}^{*}e^{\text{i}\varphi} & \Omega_{\pi}^{*}\\
-\Omega_{\pi} & -\Omega_{\sigma}e^{-\text{i}\varphi} & \Delta_{0} & 0\\
-\Omega_{\sigma}e^{-\text{i}\varphi} & \Omega_{\pi} & 0 & \Delta_{0}
\end{array}\right)
\]
in the rotating wave approximation. Here $\Omega_{\pi}=\frac{d\varepsilon_{\pi}}{\hbar}$ and $\Omega_{\sigma}=-\frac{d\varepsilon_{\sigma}}{\hbar}$ denote
the Rabi frequencies for the control and probe fields, with $d=\left\langle m_{F}=\frac{1}{2}\left|{\bf D}\cdot{\bf e}_{z}\right|m_{F}=\frac{1}{2}\right\rangle $
the $z$-component of the dipole moment of the transition $\left(F=\frac{1}{2},m_{F}=\frac{1}{2}\right)\rightarrow\left(F=\frac{1}{2},m_{F}=\frac{1}{2}\right)$,
$\Delta_{0}=\omega_{0}-\omega$ is the detuning of the fields with
respect to the DTLS natural frequency, $\omega_{0}$, while $\varepsilon_{j=\pi,\sigma}\left(y\right)=\varepsilon_{j}^{+}\left(y\right)+\varepsilon_{j}^{-}\left(y\right)e^{-2\text{i}ky}$
are the envelopes of the control and probe fields, respectively. 
 The relaxation is mainly due to %Relaxation essentially originates from 
radiative decay and residual collisions. Collisions affect Zeeman excited state and ground-excited state coherences in a different manner since the latter are sensitive to spin depolarizing collisions which leads to different rates. Therefore, in general,
the populations $\rho_{cc}$ and $\rho_{dd}$ relax with the rate $\Gamma$,
while the coherences $\rho_{ac}$, $\rho_{ad}$, $\rho_{bc}$ and $\rho_{bd}$
decay with the rate $\Gamma_{d}$ and the excited Zeeman coherence with
the rate $\Gamma_{ze}$. In the absence of non-radiative homogeneous dephasing processes, however, $\left(\Gamma_{ze},\Gamma_{d}\right)$ reduce
to $\left(\Gamma,\frac{\Gamma}{2}\right)$. 
Finally,  the  Doppler effect was implicitly neglected in the  above model% above
, which requires   the atoms  to be %are 
(sufficiently) cold.

As shown in previous works \cite{DB07,HB08a,HB08b}, the propagation of the control field is governed by the coherence $\rho_{\pi} = \rho_{ca} - \rho_{db}$, while the propagation of the probe field is  governed  %ruled 
by the coherence $\rho_{\sigma}=\rho_{cb}+\rho_{da}$
which radiates the $\sigma$-polarized light and  which obeys %whose evolution is  ruled by 
the equation
\[
\partial_{t}\rho_{\sigma}=\text{i}\Omega_{\pi}\left(\rho_{Z}-\rho_{Z}^{*}\right)+\text{i}\Omega_{\sigma}e^{-\text{i}\varphi}\left(n_{g}-n_{e}\right)-\left(\text{i}\Delta_{\pi}+\Gamma_{d}\right)\rho_{\sigma}
\]
where $\rho_{Z}=\rho_{ab}+\rho_{cd}$. The coherence $\rho_{\sigma}$
therefore  is built %builds 
up through two competing phenomena, \emph{i.e.} the
diffraction of the pump by the Zeeman coherences (first term) and
the absorption of the probe by the population (second term). In this article, we are interested in the steady state optical response, in  which %this 
case the excited state coherence is destroyed by relaxations,
$\rho_{cd}=0$, and only the ground state Zeeman coherence $\rho_{ab}$
contributes to $\rho_{\sigma}$. In the stationary regime, one finds
$\rho_{\sigma}=2\rho_{da}=2\rho_{cb}=\left(\frac{\Omega_{\sigma}e^{-\text{i}\varphi}}{\Omega_{\pi}}\right)^{*}\rho_{\pi}$
with 
\begin{align}
\rho_{\pi} & =\frac{\left(\Omega_{\pi}^{2}+\Omega_{\sigma}^{2}e^{-2\text{i}\varphi}\right)\left(\text{i}\Gamma_{d}+\Delta_{0}\right)\Omega_{\pi}^{*}}{4\Gamma_{d}\Gamma^{-1}\left|\Omega_{\pi}^{2}+\Omega_{\sigma}^{2}e^{-2\text{i}\varphi}\right|^{2}+\left(\left|\Omega_{\pi}\right|^{2}+\left|\Omega_{\sigma}\right|^{2}\right)\left(\Gamma_{d}^{2}+\Delta_{0}^{2}\right)}\label{Rhopistat}\\
\rho_{\sigma} & =\frac{\left(\Omega_{\pi}^{2}+\Omega_{\sigma}^{2}e^{-2\text{i}\varphi}\right)\left(\text{i}\Gamma_{d}+\Delta_{0}\right)\Omega_{\sigma}^{*}e^{\text{i}\varphi}}{4\Gamma_{d}\Gamma^{-1}\left|\Omega_{\pi}^{2}+\Omega_{\sigma}^{2}e^{-2\text{i}\varphi}\right|^{2}+\left(\left|\Omega_{\pi}\right|^{2}+\left|\Omega_{\sigma}\right|^{2}\right)\left(\Gamma_{d}^{2}+\Delta_{0}^{2}\right)}\label{Rhosigmastat}
\end{align}
Throughout this work, we will assume that, along the sample, the control
is much more intense than the probe, \emph{i.e. }
\begin{equation}
\left|\Omega_{\pi}\left(y\right)\right|\gg\left|\Omega_{\sigma}\left(y\right)\right|\label{ConditionPerturbation}
\end{equation}
therefore, from Eqs. (\ref{Rhopistat},\ref{Rhosigmastat}), we get
\begin{align}
\rho_{\pi} & =\frac{\text{i}\Gamma_{d}+\Delta_{0}}{4\Gamma_{d}\Gamma^{-1}\left|\Omega_{\pi}\right|^{2}+\Gamma_{d}^{2}+\Delta_{0}^{2}}\Omega_{\pi}\label{Rhopistatperturbative}\\
\rho_{\sigma} & =\frac{\text{i}\Gamma_{d}+\Delta_{0}}{4\Gamma_{d}\Gamma^{-1}\left|\Omega_{\pi}\right|^{2}+\Gamma_{d}^{2}+\Delta_{0}^{2}}\;\frac{\Omega_{\pi}^{2}}{\left|\Omega_{\pi}\right|^{2}}\Omega_{\sigma}^{*}e^{\text{i}\varphi}\label{Rhosigmastatperturbative}
\end{align}
Note that both coherences $\rho_{\sigma}$ and $\rho_{\pi}$ vanish when $\left|\Omega_{\pi}\right|\gg\sqrt{\Gamma_{d}\Gamma}$ : the saturation
of the medium by the pump  makes it  %renders the medium 
transparent for both
$\sigma$ and $\pi$ polarized fields.  This leads to effects  % Effects 
such as incoherent fluorescence% hence arise
. In the present work, however, we assume  that this regime is not reached. In other words, our treatment is non-perturbative with respect to the control field but the transparency regime is never reached.

We now define $\alpha_{0}=\frac{Nd^{2}\omega_{0}}{2c\hbar\varepsilon_{0}\Gamma_{d}}$
the field absorption coefficient at resonance, where $N$ denotes
the number of atoms per unit volume in the medium, and we introduce
$\chi_{lin}$ and $\chi_{sat}$, the linear and saturated susceptibilities
respectively, and $\chi_{eff}$ --  the meaning of which %whose signification 
will be given later,
\begin{align}
\chi_{lin} & =\frac{\frac{2\alpha_{0}\Gamma_{d}}{k}}{-\text{i}\Gamma_{d}+\Delta_{0}}\label{Chilin}\\
\chi_{sat} & =\chi_{lin}S\label{Chisat}\\
\chi_{eff} & =\chi_{sat}e^{2\text{i}\varphi}\label{Chieff}
\end{align}
with the saturation parameter $S\left(y\right)$ defined as
\begin{equation}
S\left(y\right)=\frac{1}{1+F_{0}\left|\Omega_{\pi}\left(y\right)\right|^{2}}\label{SaturationParameter}
\end{equation}
and 

\begin{equation}
F_{0}=\frac{4\Gamma_{d}\Gamma^{-1}}{\Gamma_{d}^{2}+\Delta_{0}^{2}}\label{F0Parameter}
\end{equation}
Using Eqs. (\ref{Rhopistatperturbative}-\ref{F0Parameter}), we finally
get
\begin{align}
\rho_{\pi} & =\left(\frac{k}{2\alpha_{0}\Gamma_{d}}\right)\chi_{sat}\Omega_{\pi}\label{Rhopistatperturbativefinal}\\
\rho_{\sigma} & =\left(\frac{k}{2\alpha_{0}\Gamma_{d}}\right)\chi_{eff}\frac{\Omega_{\pi}^{2}}{\left|\Omega_{\pi}\right|^{2}}\Omega_{\sigma}^{*}e^{-\text{i}\varphi}\label{Rhosigmastatperturbativefinal}
\end{align}
%{\bf Note that in the derivation above -- in particular steady-state Eqs. (\ref{Rhopistat},\ref{Rhosigmastat}) --   we implicitly assumed incomplete saturation conditions. If the medium were fully saturated it would become transparent -- coherences $\rho_{\pi}$ and $\rho_{\sigma}$ would vanish --  and neglected effects such as incoherent fluorescence should then be taken into account. This is, however, beyond the scope of the present work.}

\subsection{Propagation equations}
The control and probe fields obey the following one-dimensional Maxwell
propagation equations

\begin{align*}
\frac{\partial^{2}}{\partial y^{2}}\left(\Omega_{\pi}e^{\text{i}ky}\right)+k^{2}\left(\Omega_{\pi}e^{\text{i}ky}\right) & =-2k\alpha_{0}\Gamma_{d}\left(\rho_{\pi}e^{\text{i}ky}\right)\\
\frac{\partial^{2}}{\partial y^{2}}\left(\Omega_{\sigma}e^{\text{i}ky}e^{-\text{i}\varphi}\right)+k^{2}\left(\Omega_{\sigma}e^{\text{i}ky}e^{-\text{i}\varphi}\right) & =-2k\alpha_{0}\Gamma_{d}\left(\rho_{\sigma}e^{\text{i}ky}\right)
\end{align*}
Using Eqs. (\ref{Control},\ref{Probe}), we explicitly decompose the
Rabi frequencies into their forward and backward contributions
\begin{equation}
\Omega_{j=\pi,\sigma}\left(y\right)=\Omega_{j}^{+}\left(y\right)+\Omega_{j}^{-}\left(y\right)e^{-2\text{i}ky}\label{OmegaForwardBackward}
\end{equation}
with $\Omega_{j=\pi,\sigma}^{\pm}=\frac{d\varepsilon_{j}^{\pm}}{\hbar}$,
and we define the associated phases $\varphi_{j}^{\pm}$ by the relation
$\Omega_{j}^{\pm}=\left|\Omega_{j}^{\pm}\right|e^{-\text{i}\varphi_{j}^{\pm}}$.
We also formally expand the coherences $\rho_{j=\pi,\sigma}$ as
\begin{equation}
\rho_{j}\left(y\right)=\sum_{n=-\infty}^{+\infty}\rho_{j}^{\left(n\right)}\left(y\right)e^{2\text{i}nky}\label{RhoFloquet}
\end{equation}
When the conditions $\left|\frac{\partial^{2}}{\partial y^{2}}\Omega_{j}^{\pm}\right|\ll k\left|\frac{\partial}{\partial y}\Omega_{j}^{\pm}\right|\ll k^{2}\left|\Omega_{j}^{\pm}\right|$
are   fulfilled %met
, the slow envelope approximation can be performed. The envelopes
$\rho_{j}^{\left(n\right)}\left(y\right)$ then vary slowly on  %are then slowly varying at 
the wavelength scale. Keeping %Retaining 
only the relevant terms $\rho_{j}^{\left(n\right)}\left(y\right)$
which satisfy the phase matching condition, one finally  obtains  %gets
from
Eqs. (\ref{Rhosigmastatperturbativefinal}-\ref{RhoFloquet})
\begin{align}
\frac{\partial\Omega_{\pi}^{+}}{\partial y} & =\text{i}\alpha_{0}\Gamma_{d}\rho_{\pi}^{\left(0\right)}\label{MaxwellControl+}\\
\frac{\partial\Omega_{\pi}^{-}}{\partial y} & =-\text{i}\alpha_{0}\Gamma_{d}\rho_{\pi}^{\left(-1\right)}\label{MaxwellControl-}\\
\frac{\partial\Omega_{\sigma}^{+}}{\partial y} & =\text{i}\alpha_{0}\Gamma_{d}e^{\text{i}\varphi}\rho_{\sigma}^{\left(0\right)}\label{MaxwellProbe+}\\
\frac{\partial\Omega_{\sigma}^{-}}{\partial y} & =-\text{i}\alpha_{0}\Gamma_{d}e^{\text{i}\varphi}\rho_{\sigma}^{\left(-1\right)}\label{MaxwellProbe-}
\end{align}
These equations  could also be interpreted in terms of a four-wave mixing process  in degenerate conditions.

In the next section, we recall important results on the simpler specific
case when  the control and probe beams are both forward propagating. Then,
in Sec. \ref{TwoFields}, we  consider %shall address 
the more complex configuration  with %involving 
a stationary control field and investigate in detail the
observed phase-controlled probe reflection and transmission  by %through
analyzing the behaviour of  the associated coefficients $R=\left|\frac{\Omega_{\sigma}^{-}\left(y=0\right)}{\Omega_{\sigma}^{+}\left(y=0\right)}\right|^{2}$
and $T=\left|\frac{\Omega_{\sigma}^{+}\left(y=L\right)}{\Omega_{\sigma}^{+}\left(y=0\right)}\right|^{2}$.

\section{Situation with no backward control field\label{SingleField} }

We first address the case of forward-propagating fields inside the
medium, \emph{i.e.} $\varepsilon_{j}^{-}\left(y\right)=0$, $\Omega_{j}=\Omega_{j}^{+}$
and $\rho_{j}=\rho_{j}^{\left(0\right)}$ for $j=\pi,\sigma$. This
situation has been studied in detail in \cite{HB08b}, we shall therefore
only briefly summarize the main effects here. 

Using the relation $\frac{\Omega_{\pi}^{2}}{\left|\Omega_{\pi}\right|^{2}}\Omega_{\sigma}^{*}=e^{2\text{i}\Delta\varphi_{\sigma\pi}^{+}}\Omega_{\sigma}^{+}$
with $\Delta\varphi_{\sigma\pi}^{+}=\varphi_{\sigma}^{+}-\varphi_{\pi}^{+}$,
Eqs. (\ref{Rhopistatperturbativefinal}, \ref{Rhosigmastatperturbativefinal},
\ref{MaxwellControl+}-\ref{MaxwellProbe-}) lead to 

\begin{align}
\frac{\partial\Omega_{\pi}^{+}}{\partial y} & =\text{i}\frac{k}{2}\chi_{sat}\Omega_{\pi}^{+}\label{Maxwellcontrol+1Field}\\
\frac{\partial\left(\Omega_{\sigma}^{+}e^{-\text{i}\varphi}\right)}{\partial y} & =\text{i}\frac{k}{2}\chi_{eff}e^{2\text{i}\Delta\varphi_{\sigma\pi}^{+}}\left(\Omega_{\sigma}^{+}e^{-\text{i}\varphi}\right)\label{Maxwellprobe+1Field}
\end{align}
Provided that the dephasing between the control and probe fields,
denoted by $\Delta\varphi_{\sigma\pi}^{+}$, does not %substantially
grow  significantly along  the  propagation, $\chi_{eff}$ is seen to play the role of
an effective susceptibility for the probe in Eq. (\ref{Maxwellprobe+1Field}).
From Eqs. (\ref{Maxwellcontrol+1Field},\ref{Maxwellprobe+1Field}),
one derives the propagation equation 
\begin{equation}
\frac{\partial\Delta\varphi_{\sigma\pi}^{+}}{\partial y}=k\sin\left(\Delta\varphi_{\sigma\pi}^{+}-\varphi\right)\left[\chi'_{sat}\sin\left(\Delta\varphi_{\sigma\pi}^{+}-\varphi\right)-\chi"_{sat}\cos\left(\Delta\varphi_{\sigma\pi}^{+}-\varphi\right)\right]\label{PropagationDphi}
\end{equation}
where $\chi'_{sat}$ and $\chi"_{sat}$ denote the real and imaginary
parts of the saturated susceptibility, respectively. We introduce
the phase of the linear susceptibility, $\varphi_{L}$, defined by
$\chi_{lin}=\left|\chi_{lin}\right|e^{\text{i}\varphi_{L}}$, which
verifies 
\begin{equation}
\tan\varphi_{L}=\frac{\Gamma_{d}}{\Delta_{0}}\label{TanphiL}
\end{equation}

The integration Eq. (\ref{PropagationDphi}) leads to $\tan\Delta\varphi_{\sigma\pi}^{+}=\frac{\tan\varphi_{L}}{\left(\frac{\tan\varphi_{L}}{\tan\varphi}+1\right)e^{-2\alpha_{0}y\sin^{2}\varphi_{L}}-1}$.
When $\left|\Omega_{\pi}\left(y=0\right)\right|\ll\sqrt{\Gamma_{d}\Gamma}$
and $\alpha_{0}L\sin^{2}\varphi_{L}\ll1$, then $\left|\Omega_{\pi}^{+}\left(y\right)\right|\approx\left|\Omega_{\pi}^{+}\left(y=0\right)\right|$,
$S\approx1$ and $\Delta\varphi_{\sigma\pi}^{+}\approx0$, see \cite{HB08b}.
The sample therefore becomes a linear medium for the probe field whose
susceptibility, $\chi_{eff}$, is related to the true linear susceptibility
 by %through 
$\chi_{eff}=\chi_{lin}e^{2\text{i}\varphi}$. The latter relation
proves the existence of the phase control of the linear response of
the medium,  the %whose 
physical interpretation  of which  was described in \cite{HB08b}.
It results from interference effects between the quantum paths which
contribute to the coherence $\rho_{\sigma}$ responsible for the $\sigma$-polarized
radiated field. 

 As %If 
the control field intensity is increased, saturation effects occur
and the validity condition of  the  parametric approximation  becomes %is 
less restrictive.
From Eq. (\ref{Maxwellcontrol+1Field}), one gets 
\[
\frac{\partial\left|\Omega_{\pi}^{+}\right|}{\partial y}=-\alpha_{0}S\sin^{2}\varphi_{L}\left|\Omega_{\pi}^{+}\right|
\]
and provided that $\alpha_{0}LS\left(y=0\right)\sin^{2}\varphi_{L}\ll1$,
 the intensity of the  control field %intensity 
remains unaffected along propagation. If the
optical depth is increased so that the inequality $\alpha_{0}LS\left(y=0\right)\sin^{2}\varphi_{L}\ll1$
no longer holds, the  change in  %modification of 
the control field cannot be neglected
and additional dephasing  must %has to 
be taken into account in the propagation
equation of the probe. New features appear and are detailed in \cite{HB08b}. 

\section{Situation of a stationary control field\label{TwoFields}}
In this section, we investigate the effects which appear when the
intensity of the control field is not uniform but spatially modulated
because of the backward component. In this situation, the probe field
experiences both reflection and transmission. The  effect  %impact 
of the  change in %modification of 
the optical response of the sample and the  possibility %eventuality 
of phase control can be evaluated by calculating the transmission and reflection
coefficients for the probe field. 

\subsection{Propagation equations in the slow envelope approximation }

In the slow envelope approximation, the set of Eqs. (\ref{MaxwellControl+}-\ref{MaxwellProbe-})
can be rewritten by extracting an analytical expression for the components
$\rho_{j}^{\left(0\right)}\left(y\right)$ and $\rho_{j}^{\left(-1\right)}\left(y\right)$. 

We introduce the parameter $r\left(y\right)$ and  the phase $\varphi_{r}$
defined by $\Omega_{\pi}^{-}\left(y\right)=r\left(y\right)\Omega_{\pi}^{+}\left(y\right)$
and $r=\left|r\right|e^{-\text{i}\varphi_{r}}$, so that $\varphi_{r}=\varphi_{\pi}^{-}-\varphi_{\pi}^{+}$.
When the control field is unaffected by the propagation, $r\left(y\right)$
is constant. Note that, for Eq. (\ref{ConditionPerturbation}) to hold,
the control field should not vanish anywhere, including the spatial
nodes, therefore $r$ should be different from $1$. From Eqs. (\ref{Rhopistatperturbativefinal},\ref{Rhosigmastatperturbativefinal},\ref{RhoFloquet}),
we see that the analytic expressions of $\rho_{j}^{\left(0\right)}\left(y\right)$
and $\rho_{j}^{\left(-1\right)}\left(y\right)$ can be evaluated as
soon as the series expansion of the saturation parameter $S\left(y\right)$
and $\frac{1}{\left|\Omega_{\pi}\left(y\right)\right|^{2}}$  is %are 
determined.
The saturation parameter $S$ given by Eq. (\ref{SaturationParameter})
 can %may 
be expressed as
\begin{equation}
S\left(y\right)=\frac{b\left(y\right)}{1+a\left(y\right)\cos\left[2ky+\varphi_{r}\left(y\right)\right]}\label{SaturationParameter2Fields}
\end{equation}
with $a\left(y\right)=\frac{2F_{0}\left|\Omega_{\pi}^{+}\left(y\right)\right|^{2}\left|r\left(y\right)\right|}{1+F_{0}\left|\Omega_{\pi}^{+}\left(y\right)\right|^{2}\left(1+\left|r\left(y\right)\right|^{2}\right)}$
and $b\left(y\right)=\frac{1}{1+F_{0}\left|\Omega_{\pi}^{+}\left(y\right)\right|^{2}\left(1+\left|r\left(y\right)\right|^{2}\right)}$.
It can also be formally expanded as
\begin{equation}
S\left(y\right)=\sum_{n=-\infty}^{+\infty}c_{n}\left(y\right)e^{2\text{i}nky}\label{ExpansionS}
\end{equation}
where the $c_{n}$ coefficients %are slowly varying 
 vary slowly on %at 
the spatial
wavelength scale $\frac{2\pi}{k}$ and can be approximated as
$c_{n}\left(y\right)\approx\frac{k}{\pi}\int_{y}^{y+\frac{\pi}{k}}S\left(y'\right)e^{-2\text{i}nky'}\text{d}y'$.
For $n\geq0$, using Eq. (\ref{SaturationParameter2Fields}) and \cite{GR14},
we get $c_{n}\left(y\right)\approx c_{0}\eta^{n}e^{\text{i}n\varphi_{r}\left(y\right)}$,
with $c_{0}=\frac{b\left(y\right)}{\sqrt{1-a^{2}\left(y\right)}}$
and $\eta=\frac{\sqrt{1-a^{2}\left(y\right)}-1}{a\left(y\right)}$.
 Furthermore,   %Moreover, 
for $n\leq0$, the reality of parameter $S$ implies that
$c_{n}=c_{-n}^{*}$.

On the other hand, using Eq. (\ref{OmegaForwardBackward}), one  can  %may
put the parameter $\frac{1}{\left|\Omega_{\pi}\left(y\right)\right|^{2}}$
 in the form
\begin{equation}
\frac{1}{\left|\Omega_{\pi}\left(y\right)\right|^{2}}=\frac{b'\left(y\right)}{1+a'\left(y\right)\cos\left[2ky+\varphi_{r}\left(y\right)\right]}\label{InverseOmegaPiSquared}
\end{equation}
with $a'\left(y\right)=\frac{2\left|r\left(y\right)\right|}{\left|\Omega_{\pi}^{+}\left(y\right)\right|^{2}\left(1+\left|r\left(y\right)\right|^{2}\right)}$
and $b'\left(y\right)=\frac{1}{\left|\Omega_{\pi}^{+}\left(y\right)\right|^{2}\left(1+\left|r\left(y\right)\right|^{2}\right)}$.
It can also be formally expanded as $\frac{1}{\left|\Omega_{\pi}\left(y\right)\right|^{2}}=\sum_{n=-\infty}^{+\infty}c_{n}'\left(y\right)e^{2\text{i}nky}$
where the $c_{n}'$ coefficients  vary slowly on  %are slowly varying at 
the wavelength
scale and can be expressed as $c_{n}'\left(y\right)\approx\frac{k}{\pi}\int_{y}^{y+\frac{\pi}{k}}\frac{1}{\left|\Omega_{\pi}\left(y'\right)\right|^{2}}e^{-2\text{i}nky'}\text{d}y'$.
Using \cite{GR14}, we get $c'_{n\geq0}\left(y\right)\approx c'_{0}\eta'^{n}e^{\text{i}n\varphi_{r}\left(y\right)}$
with $c'_{0}=\frac{b'\left(y\right)}{\sqrt{1-a'^{2}\left(y\right)}}$
and $\eta'=\frac{\sqrt{1-a'^{2}\left(y\right)}-1}{a'\left(y\right)}$.
 Furthermore,  %Moreover, 
the reality of $\frac{1}{\left|\Omega_{\pi}\right|^{2}}$
implies that $c'_{n\leq0}=\left(c'_{-n}\right)^{*}$.

Using Eqs. (\ref{Rhopistatperturbativefinal},\ref{Rhosigmastatperturbativefinal},\ref{RhoFloquet},\ref{SaturationParameter2Fields},\ref{ExpansionS},\ref{InverseOmegaPiSquared}),
we can finally express the components of the coherences which appear
in the propagation equations for the control fields
\begin{align}
\rho_{\pi}^{\left(0\right)}\left(y\right) & =\frac{\text{i}}{\Gamma_{d}}\left[\overline{c}_{0}\left(y\right)\Omega_{\pi}^{+}\left(y\right)+\overline{c}_{1}\left(y\right)\Omega_{\pi}^{-}\left(y\right)\right]\label{RhoControl0}\\
\rho_{\pi}^{\left(-1\right)}\left(y\right) & =\frac{\text{i}}{\Gamma_{d}}\left[\overline{c}_{-1}\left(y\right)\Omega_{\pi}^{+}\left(y\right)+\overline{c}_{0}\left(y\right)\Omega_{\pi}^{-}\left(y\right)\right]\label{RhoControl-1}
\end{align}
with $\overline{c}_{i}\left(y\right)=\left(\frac{\Gamma_{d}}{\Gamma_{d}+\text{i}\Delta_{0}}\right)c_{i}\left(y\right)$,
and in the propagation equations for the probe fields
\begin{align}
\rho_{\sigma}^{\left(0\right)}\left(y\right) & =\frac{\text{i}}{\Gamma_{d}}e^{\text{i}\varphi}\left[c_{+}\left(y\right)\Omega_{\sigma}^{+*}\left(y\right)+c_{+-}\left(y\right)\Omega_{\sigma}^{-*}\left(y\right)\right]\label{Rhosigma0}\\
\rho_{\sigma}^{\left(-1\right)}\left(y\right) & =\frac{\text{i}}{\Gamma_{d}}e^{\text{i}\varphi}\left[c_{+-}\left(y\right)\Omega_{\sigma}^{+*}\left(y\right)+c_{-}\left(y\right)\Omega_{\sigma}^{-*}\left(y\right)\right]\label{Rhosigma-1}
\end{align}
with
\begin{align}
c_{+} & =\beta\left[1+2\gamma\left|r\right|+\delta\left|r\right|^{2}\right]\label{Coefficientc+}\\
c_{+-} & =\beta r\left[2+\frac{\gamma}{\left|r\right|}+\gamma\left|r\right|\right]\nonumber \\
c_{-} & =\beta r^{2}\left[1+\frac{2\gamma}{\left|r\right|}+\frac{\delta}{\left|r\right|^{2}}\right]\nonumber 
\end{align}
and
\begin{align}
\beta & =\frac{c_{0}e^{-2\text{i}\varphi_{\pi}^{+}}}{\left(1-\eta\eta'\right)\left|1-\left|r\right|^{2}\right|}\left(\frac{\Gamma_{d}}{\Gamma_{d}+\text{i}\Delta_{0}}\right)\label{Coefficientbeta}\\
\gamma & =\eta+\eta'\label{Coefficientgamma}\\
\delta & =\eta^{2}+\left(\eta'\right)^{2}+\eta\eta'-\left(\eta\eta'\right)^{2}\label{Coefficientdelta}
\end{align}
Injecting Eqs. (\ref{RhoControl0}-\ref{Rhosigma-1}) into Eqs. (\ref{MaxwellControl+},\ref{MaxwellControl-})
and Eqs. (\ref{MaxwellProbe+},\ref{MaxwellProbe-}), we get
\begin{align}
\frac{\partial\Omega_{\pi}^{+}}{\partial\left(\frac{y}{L}\right)}\left(y\right) & =-\alpha_{0}L\left[\overline{c}_{0}\left(y\right)\Omega_{\pi}^{+}\left(y\right)+\overline{c}_{1}\left(y\right)\Omega_{\pi}^{-}\left(y\right)\right]\label{MaxwellControl+2Fields}\\
\frac{\partial\Omega_{\pi}^{-}}{\partial\left(\frac{y}{L}\right)}\left(y\right) & =\alpha_{0}L\left[\overline{c}_{-1}\left(y\right)\Omega_{\pi}^{+}\left(y\right)+\overline{c}_{0}\left(y\right)\Omega_{\pi}^{-}\left(y\right)\right]\label{MaxwellControl-2Fields}
\end{align}
and
\begin{align}
\frac{\partial\Omega_{\sigma}^{+}}{\partial\left(\frac{y}{L}\right)}\left(y\right) & =-\alpha_{0}Le^{2\text{i}\varphi}\left[c_{+}\left(y\right)\Omega_{\sigma}^{+*}\left(y\right)+c_{+-}\left(y\right)\Omega_{\sigma}^{-*}\left(y\right)\right]\label{MaxwellProbe+2Fields}\\
\frac{\partial\Omega_{\sigma}^{-}}{\partial\left(\frac{y}{L}\right)}\left(y\right) & =\alpha_{0}Le^{2\text{i}\varphi}\left[c_{+-}\left(y\right)\Omega_{\sigma}^{+*}\left(y\right)+c_{-}\left(y\right)\Omega_{\sigma}^{-*}\left(y\right)\right]\label{MaxwellProbe-2Fields}
\end{align}
The forward and backward components of both the probe and control
fields are now coupled through the coefficients $\left(\overline{c}_{-1},\overline{c}_{1}\right)$
for the control and $c_{+-}$ coefficient for the probe. In particular,
Eqs. (\ref{MaxwellProbe+2Fields},\ref{MaxwellProbe-2Fields}) clearly
show that the probe field propagation may be controlled by the phase.
This property is further investigated in Sec. \ref{NumericalResults},
through numerically solving the above coupled equations.

\subsection{Parametric regime\label{SmallOD}}

The probe field propagation depends on the control field intensity
and the optical depth. For moderate optical depths, \emph{i.e.} $\left|\alpha_{0}L\overline{c}_{j=0,\pm1}\right| \ll 1$,
the control field is weakly affected by  the propagation and the envelopes
can be considered uniform, \emph{i.e.} $\Omega_{\pi}^{+}\left(y\right)\approx\Omega_{\pi}^{+}\left(0\right)$,
$\Omega_{\pi}^{-}\left(y\right)\approx\Omega_{\pi}^{-}\left(L\right)$
and $\varphi_{\pi}^{\pm}\approx0$. The coefficients $c_{+},c_{-},c_{+-}$
are thus constant along the sample and the phases $\varphi_{\pi}^{+}$
and $\varphi_{r}$ are neglected. The system of Eqs. (\ref{MaxwellProbe+2Fields},\ref{MaxwellProbe-2Fields})
turns into a set of differential equations with constant coefficients.
Moreover, for optical depths such as $\left|\alpha_{0}Lc_{j=+,-,+-}\right|\ll1$,
one  can %may 
retain the most significant terms in the integration of the
system Eqs. (\ref{MaxwellProbe+2Fields},\ref{MaxwellProbe-2Fields})
which leads to 
\begin{align*}
\Omega_{\sigma}^{+}\left(y\right)-\Omega_{\sigma}^{+}\left(0\right) & \approx-\alpha_{0}ye^{2\text{i}\varphi}\left[c_{+}\Omega_{\sigma}^{+*}\left(0\right)+c_{+-}\Omega_{\sigma}^{-*}\left(0\right)\right]\\
\Omega_{\sigma}^{-}\left(y\right)-\Omega_{\sigma}^{-}\left(L\right) & \approx\alpha_{0}\left(y-L\right)e^{2\text{i}\varphi}\left[c_{+-}\Omega_{\sigma}^{+*}\left(L\right)+c_{-}\Omega_{\sigma}^{-*}\left(L\right)\right]
\end{align*}
Taking into account the boundary condition $\left\{ \Omega_{\sigma}^{-}\left(L\right)=0\;;\;\Omega_{\sigma}^{+*}\left(0\right)=\Omega_{\sigma}^{+}\left(0\right)\right\} $,
one readily obtains

\begin{align}
\Omega_{\sigma}^{+}\left(y\right) & \approx\Omega_{\sigma}^{+}\left(0\right)\left[1-\alpha_{0}ye^{2\text{i}\varphi}c_{+}\right]\label{Omegasigma+final}\\
\Omega_{\sigma}^{-}\left(y\right) & \approx\Omega_{\sigma}^{+}\left(0\right)\alpha_{0}\left(y-L\right)e^{2\text{i}\varphi}\left[c_{+-}-\alpha_{0}\left(y-L\right)e^{-2\text{i}\varphi}c_{+-}c_{+}^{*}\right]\label{Omegasigma-final}
\end{align}
From Eqs. (\ref{Coefficientc+},\ref{Coefficientbeta}-\ref{MaxwellControl-2Fields}),
we get $\arg\left[c_{+}\left(0\right)\right]=\varphi_{L}-\frac{\pi}{2}$,
where $\varphi_{L}$ is defined in Eq. (\ref{TanphiL}).  The  transmission
and reflection coefficients can then be approximated as 
\begin{align}
T & \approx1-2\alpha_{0}L\left|c_{+}\right|\sin\left(2\varphi+\varphi_{L}\right)\label{TransmissionApprox}\\
R & \approx\left|\alpha_{0}Lc_{+-}\right|^{2}\left[1-2\alpha_{0}L\left|c_{+}\right|\sin\left(2\varphi+\varphi_{L}\right)\right]\label{ReflectionApprox}
\end{align}
These expressions show the existence of \emph{phase-controlled} reflection
and transmission in the limit of parametric interaction regime where the control field is slightly affected by propagation and
the probe phase variations along propagation are weak. The coefficients
can be modified by adjusting a versatile and fine-tuning experimental parameter, $\varphi$, the additional shift $\varphi_{L}$ being determined by the detuning $\Delta_{0}$. The phase dependence is not altered by other fluctuating
parameters such as density, temperature, driving field intensity etc., and the phase thus constitutes a
genuine control parameter.
For $0\leq2\varphi+\varphi_{L}\leq\pi$,
the transmission factor $T$ is less  than or equal to $1$, while it becomes  greater %larger 
than $1$ for $\pi\leq2\varphi+\varphi_{L}\leq2\pi$. This is  consistent %in agreement 
with the  absorbing or amplifying %absorptive or amplificative 
nature of the sample as described by its effective linear susceptibility, given by Eq. (\ref{Chieff}). The reflection factor is associated with the backward component of the control field that is generated  by %through the 
coupling with the forward part of the
probe. Thus, the reflection factor depends on the second (or higher) power of the optical depth and is small %with respect 
 compared to the transmission factor.

At this point, let us make an important remark. The phase dependence   was  %could 
certainly   to  be expected from the beginning since the polarization of the total field injected into the atomic medium is elliptical and depends on $\varphi$. It is therefore natural that the intensities of the probe field reflected/transmitted by the medium accordingly depend on $\varphi$ \cite{KW17}. \emph{However, the phase-sensitivity turns to be phase-control only in the parametric regime as identified and discussed above.} This gives rise to a small reflection factor but surprisingly a very efficient amplification and modulation of the transmission factor. The mixing of the forward radiated field with the incident one, Eq. (\ref{Omegasigma+final}), is responsible for this efficiency and is reminiscent of   the  homodyne detection process. These results and the exact quantitative expressions of the necessary conditions constitute an important part of the present work.

\section{Results and discussion\label{NumericalResults} }

In this section, we present both the numerical results obtained  from  %through
the numerical simulation of Eqs. (\ref{MaxwellControl+2Fields}-\ref{MaxwellProbe-2Fields})
and   the  analytical approximation, Eqs. (\ref{TransmissionApprox},\ref{ReflectionApprox}),
and discuss their physical interpretation. 

\begin{figure}
\begin{centering}
\includegraphics[width= \textwidth ]{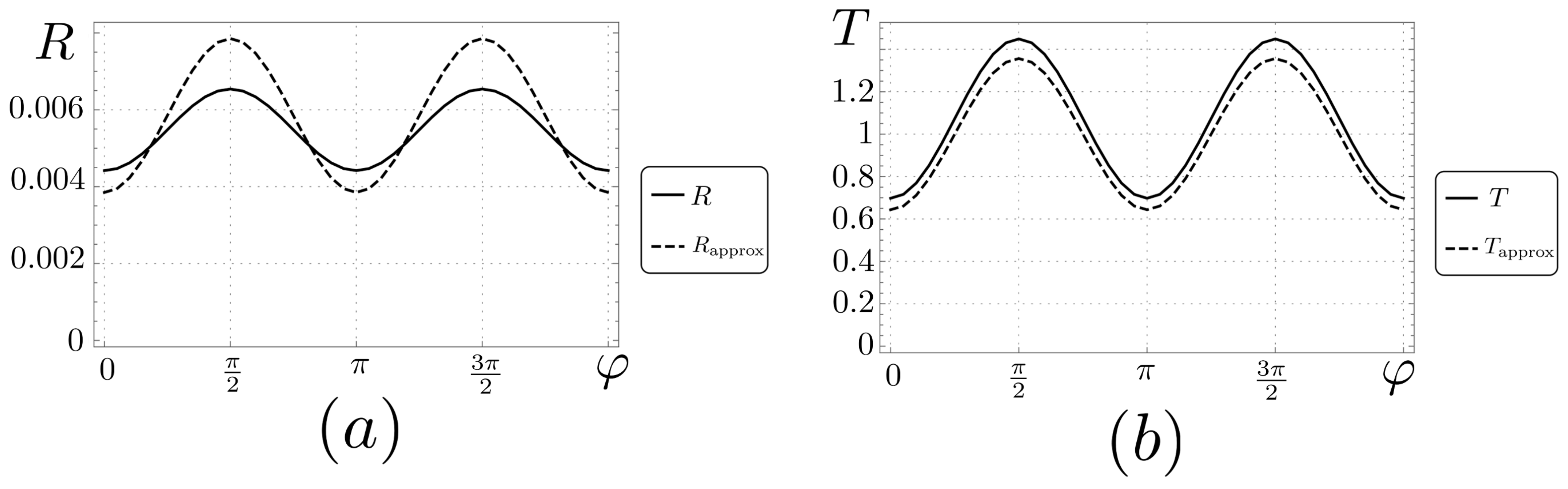}
\par\end{centering}
\caption{(a) Reflection and (b) Transmission factors as functions of the relative
phase $\varphi$ at resonance $\left(\Delta_{0}=0\right)$ for a probe
field propagating in an optically thin sample (optical depth $\alpha_{0}L=0.3$)
driven by a control field with Rabi amplitudes $\frac{\Omega_{\pi}^{+}\left(0\right)}{\Gamma_{d}}=0.4$
and $\frac{\Omega_{\pi}^{-}\left(L\right)}{\Gamma_{d}}=0.16$.
%Blue solid (orange dashed) 
%Thick solid (thin dot-
 Full  (dashed) line plots are obtained through numerical
simulation (analytic approximation). \label{Fig3}}
\end{figure}

In Fig.\ref{Fig3} we plotted %the variations of 
 (a) the reflection,
$R$, and (b) transmission, $T$, coefficients as functions of the
relative phase $\varphi$, for resonant probe and control fields $\left(\Delta_{0}=0\right)$
and small optical depth $\alpha_{0}L=0.3$.   The  forward/backward control
field Rabi frequencies at the   input/output  %entrance/exit 
of the sample are set
to $\frac{\Omega_{\pi}^{+}\left(0\right)}{\Gamma_{d}}=0.4$ and $\frac{\Omega_{\pi}^{-}\left(L\right)}{\Gamma_{d}}=0.16$,
respectively. The curves in solid and dashed lines correspond to numerical
simulations and approximate analytical expressions, Eqs. (\ref{TransmissionApprox},\ref{ReflectionApprox}),
respectively. We observe that both $R$ and $T$ oscillate with a
period of $\pi$ as expected from the $e^{2\text{i}\varphi}$ dependence
of the effective susceptibility, Eq. (\ref{Chieff}). As already pointed
out in Subsec. \ref{SmallOD}, $T$   can  %may 
be   greater   %larger 
or %smaller 
  less   than
$1$, depending   on   whether the sample is   amplifying or absorbing.  % amplificative or absorptive.
We note that $T$ is much larger than $R$ : the transmitted field
indeed results from the interference of the incident and radiated
probe fields, while, in contrast, the reflected field results from the backward radiated field only. Moreover, we observe discrepancies between the numerical simulations and the simple perturbative analytical model. 

\begin{figure}
\begin{centering}
\includegraphics[width= 0.6 \textwidth]{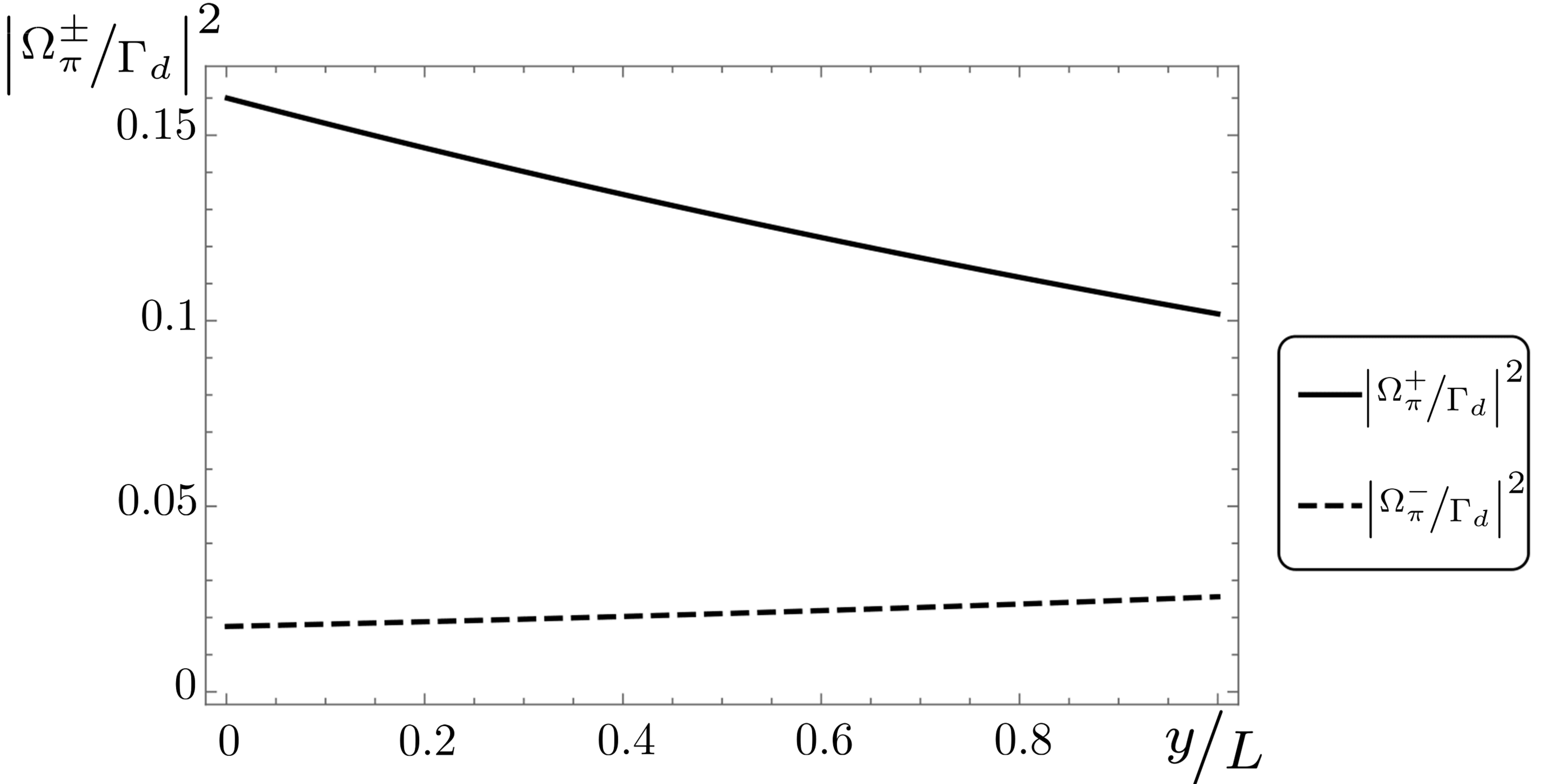}
\par\end{centering}
\caption{Control field %(a) 
intensities $\left|\frac{\Omega_{\pi}^{\pm}}{\Gamma_{d}}\right|^{2}$
%and (b) phases $\varphi_{\pi}^{\pm}$ 
as functions of the (normalized)
propagation distance $\frac{y}{L}$ for $\Delta_{0}=0$, $\alpha_{0}L=0.3$,
$\frac{\Omega_{\pi}^{+}\left(0\right)}{\Gamma_{d}}=0.4$ and $\frac{\Omega_{\pi}^{-}\left(L\right)}{\Gamma_{d}}=0.16$.
\label{Fig4}}
\end{figure}

In Fig. \ref{Fig4} %are 
  we  plotted %(a) 
the (normalized) backward and forward-propagating control field intensities $\left|\frac{\Omega_{\pi}^{\pm}\left(y\right)}{\Gamma_{d}}\right|^{2}$
%and (b) phases $\varphi_{\pi}^{\pm}$, respectively, 
as functions of the (normalized) position $\frac{y}{L}$, for the set of parameters
$\Delta_{0}=0$, $\alpha_{0}L=0.3$, $\frac{\Omega_{\pi}^{+}\left(0\right)}{\Gamma_{d}}=0.4$,
$\frac{\Omega_{\pi}^{-}\left(L\right)}{\Gamma_{d}}=0.16$.   The intensity of the forward-propagating field  %Forward-propagating field intensity 
is attenuated by $37.5\%$ along its propagation,
from the % input 
value  of  $0.16$ at the   input   %entrance 
to   about  %approximately 
$0.1$ at the  output  %exit 
of the sample, while the   intensity of the  
backward-propagating control
field %intensity 
is reduced by $22\%$, from the  %input 
 value   of  $0.0256$
at the   output  %exit 
to $0.02$ at the   input of the sample. 
%entrance. 
The phases accumulated along
  the  propagation are found to be small in all cases (of the order of $10^{-10})$. This can be understood from Eqs. (\ref{MaxwellControl+2Fields},\ref{MaxwellControl-2Fields}).
At moderate optical depth, the dominant contribution due to the diagonal
terms involves the $\overline{c}_{0}$ coefficient which is real at
resonance. The control field components therefore experience pure
absorption without dispersion, the accumulated phase is thus negligible.
For the lowest-order solution of Eqs. (\ref{MaxwellProbe+2Fields},\ref{MaxwellProbe-2Fields}),
namely Eqs. (\ref{Omegasigma+final},\ref{Omegasigma-final}), to hold,
the condition $\left|\alpha_{0}Lc_{j}\right|\ll1$ must be checked
: we find the values $0.192$, $0.005$, $0.054$ for $\left|\alpha_{0}Lc_{j=+,-,+-}\right|$,
respectively, which ensures the validity of the analytical model presented
in Subsec. \ref{SmallOD}, although the control field can be considered
only roughly unaltered along propagation. This simple model captures
the essential physical phenomena at work and clearly   demonstrates  %exhibits 
the accuracy of phase control of $R$ and $T$.

\begin{figure}
\begin{centering}
\includegraphics[width= \textwidth]{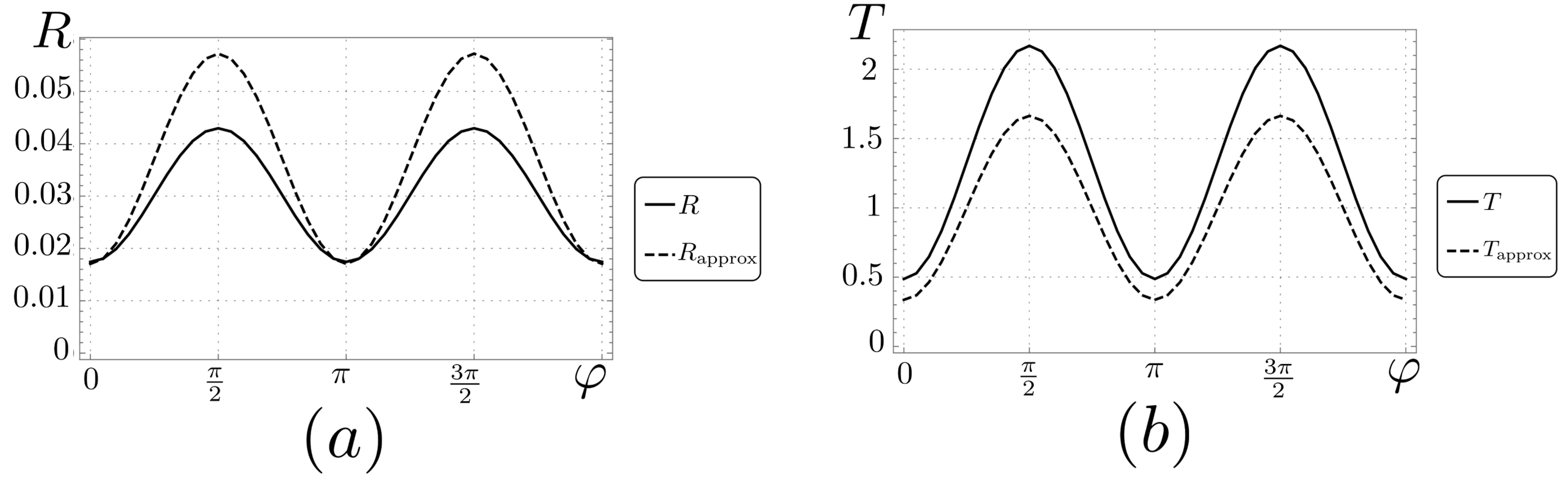}
\par\end{centering}
\caption{(a) Reflection and (b) Transmission factors as functions of the relative
phase $\varphi$ for $\Delta_{0}=0$, $\alpha_{0}L=0.6$, $\frac{\Omega_{\pi}^{+}\left(0\right)}{\Gamma_{d}}=0.4$
and $\frac{\Omega_{\pi}^{-}\left(L\right)}{\Gamma_{d}}=0.16$.
%Blue solid (resp. orange dashed) 
%Thick solid (thin dot-dashed) 
 Full (dashed)  line plots are obtained through numerical
simulation (resp. analytic approximation). \label{Fig5}}
\end{figure}

\begin{figure}
\begin{centering}
\includegraphics[width=0.6 \textwidth]{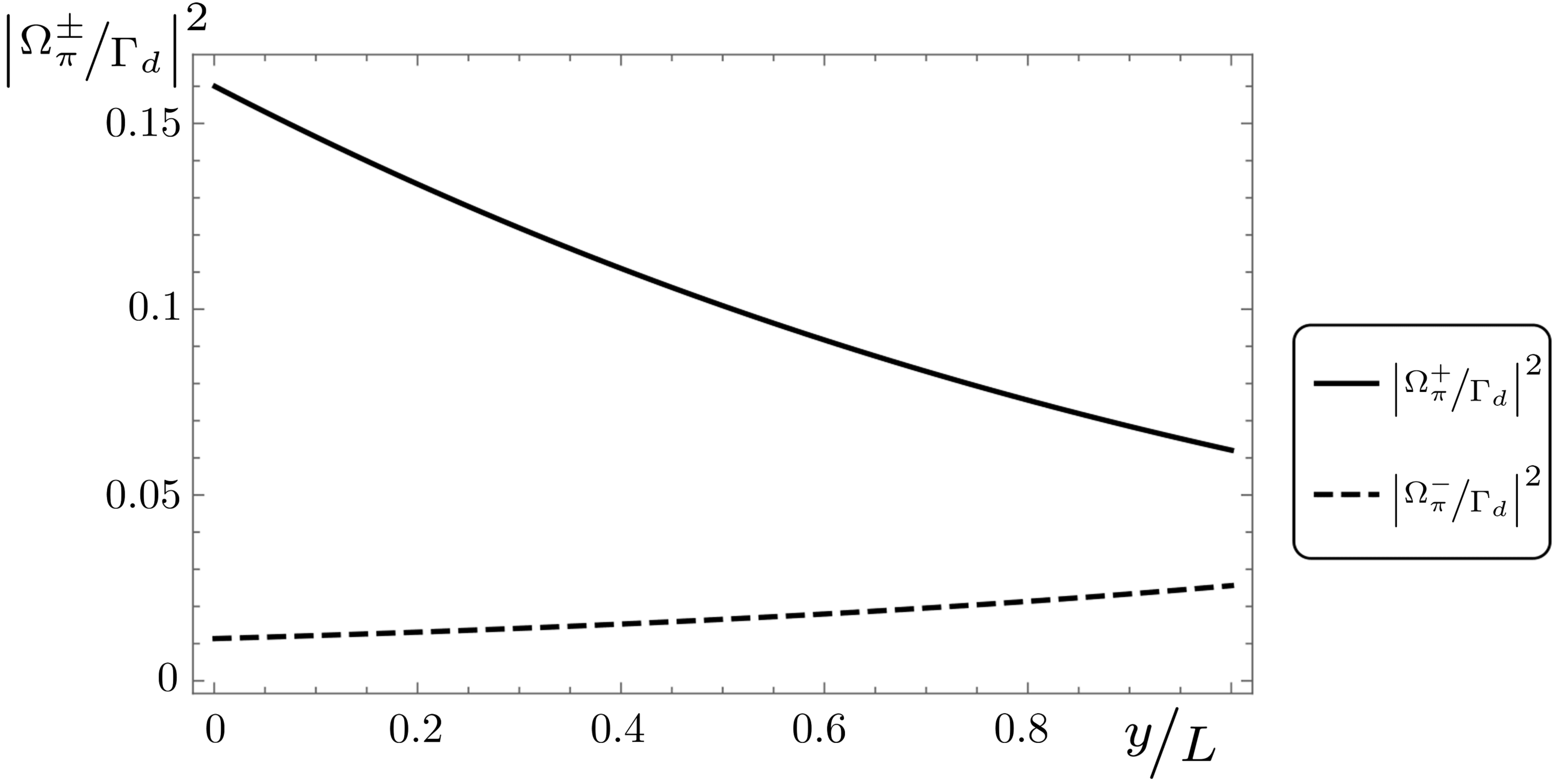}
\par\end{centering}
\caption{Control field %(a) 
intensities $\left|\frac{\Omega_{\pi}^{\pm}}{\Gamma_{d}}\right|^{2}$
%and (b) phases $\varphi_{\pi}^{\pm}$ 
as functions of the (normalized)
propagation distance $\frac{y}{L}$ for $\Delta_{0}=0$, $\alpha_{0}L=0.6$,
$\frac{\Omega_{\pi}^{+}\left(0\right)}{\Gamma_{d}}=0.4$ and $\frac{\Omega_{\pi}^{-}\left(L\right)}{\Gamma_{d}}=0.16$.
\label{Fig6}}
\end{figure}

Figs \ref{Fig5} and \ref{Fig6} correspond to the same situations
as considered in Figs \ref{Fig3} and \ref{Fig4},   but  %though 
with a higher
optical depth $\alpha_{0}L=0.6$. Both the reflection and transmission
coefficients are significantly enhanced in Fig. \ref{Fig5} compared
to Fig. \ref{Fig3} : now, $T$ oscillates between $0.48$ and $2.17$
-- versus $0.7$ and $1.45$ for $\alpha_{0}L=0.3$ -- and $R$
between $0.017$ and $0.043$ -- versus $0.0043$ and $0.0063$ for
$\alpha_{0}L=0.3$. Here, although the parametric approximation is
no longer valid, both $R$ and $T$ still depend on the phase and
are  significantly enhanced.  %substantially amplified. 
The fit %ting 
by the analytical model,
Eqs. (\ref{TransmissionApprox},\ref{ReflectionApprox}), is only qualitative
for two reasons. First, as shown in Fig. \ref{Fig6}, the increased
optical depth leads to stronger absorption of the control field along
propagation. The  intensity of the   forward-, resp. backward-, propagating component
%intensity 
is indeed reduced by $60\%$ ($37\%$ in the case of Fig.
\ref{Fig4} with $\alpha_{0}L=0.3$), resp. $56\%$ ($22\%$ in Fig.
\ref{Fig4}). The parametric approximation, $\Omega_{\pi}^{+}\left(y\right)\approx\Omega_{\pi}^{+}\left(0\right)\;;\;$$\Omega_{\pi}^{-}\left(y\right)\approx\Omega_{\pi}^{-}\left(L\right)$,
is thus rough. Secondly, the coefficients $\left|\alpha_{0}Lc_{j=+,-,+-}\right|$
are  approximately   magnified %approximatively 
by a factor  of   two  compared  %with respect 
to the
case where $\alpha_{0}L=0.3$ (e.g. $\left|\alpha_{0}Lc_{j}\right|\approx0.41,0.006,0.09$
for $j=+,-,+-$, respectively) making the low order perturbation solution
less valid. Note that in this case the dispersive component of the
optical response vanishes at resonance and therefore the phase remains
essentially constant.

\begin{figure}
\begin{centering}
\includegraphics[width=\textwidth]{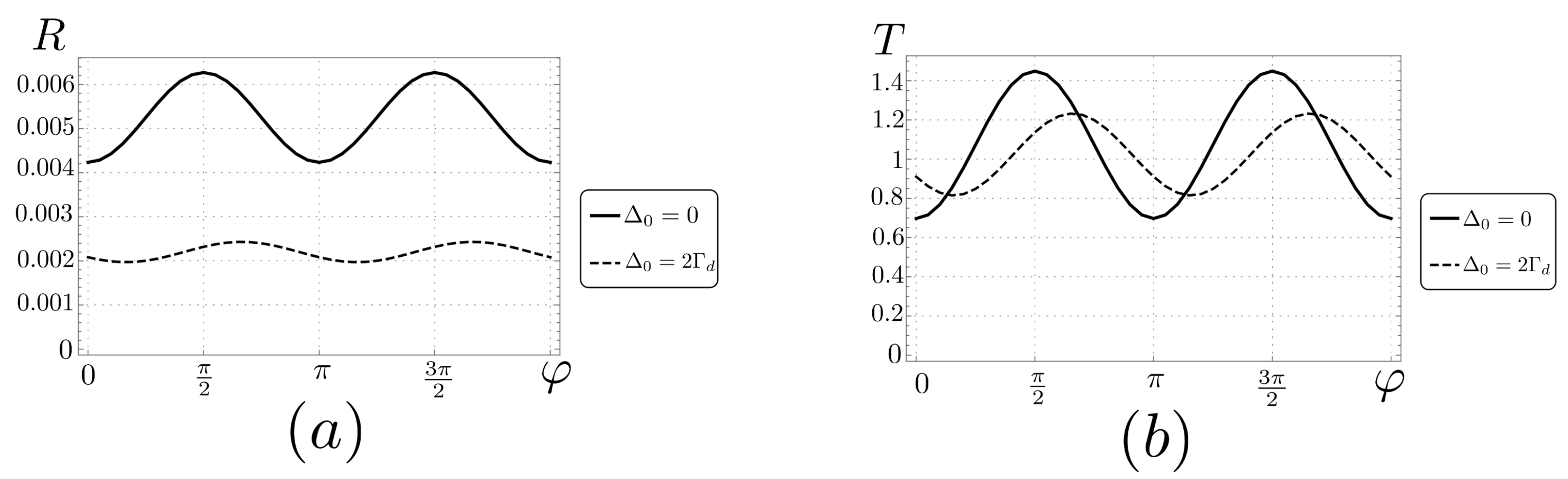}
\par\end{centering}
\caption{Effect of detuning. (a) Reflection and (b) Transmission factors plotted
as functions of the relative phase $\varphi$ for $\alpha_{0}L=0.3$,
$\frac{\Omega_{\pi}^{+}\left(0\right)}{\Gamma_{d}}=0.4$, $\frac{\Omega_{\pi}^{-}\left(L\right)}{\Gamma_{d}}=0.16$
and $\frac{\Delta_{0}}{\Gamma_{d}}=0\text{ (full line)},2\;\text{(dashed line)}$.
\label{Fig7}}
\end{figure}

In Fig. \ref{Fig7}  we  %are 
plotted the (a) reflection and (b) transmission
coefficients as functions of the relative phase $\varphi$, in the
resonant case $\Delta_{0}=0$ (solid line) and in   the non-resonant case
$\frac{\Delta_{0}}{\Gamma_{d}}=2$ (dashed line), the other parameters
being $\alpha_{0}L=0.3$ and $\frac{\Omega_{\pi}^{+}\left(0\right)}{\Gamma_{d}}=0.4$,
$\frac{\Omega_{\pi}^{-}\left(L\right)}{\Gamma_{d}}=0.16$.  As   %When
the detuning increases, the modulus of the effective susceptibility
$\chi_{eff}$, Eq. (\ref{Chieff}), decreases and so does the global
coupling of the probe  to   %with 
the atomic medium. $R$ and $T$ therefore
tend to decrease, which is confirmed by the results of   the   numerical simulations.
An important feature, however, is the dephasing which appears on $R$
and $T$ plots when %modifying 
the detuning  is modified. The expression  for   %of 
the phase shift predicted by the analytical model, Eqs. (\ref{TransmissionApprox},\ref{ReflectionApprox}),
is $\varphi_{L}$ with $\tan\varphi_{L}=\frac{\Gamma_{d}}{\Delta_{0}}$.
When $\frac{\Gamma_{d}}{\Delta_{0}}=0.5$, $\varphi_{L}\approx0.15\pi$
which is very close to the value  of   $0.16\pi$ obtained  from   %via 
numerical simulations. 

\begin{figure}
\begin{centering}
\includegraphics[width=\textwidth]{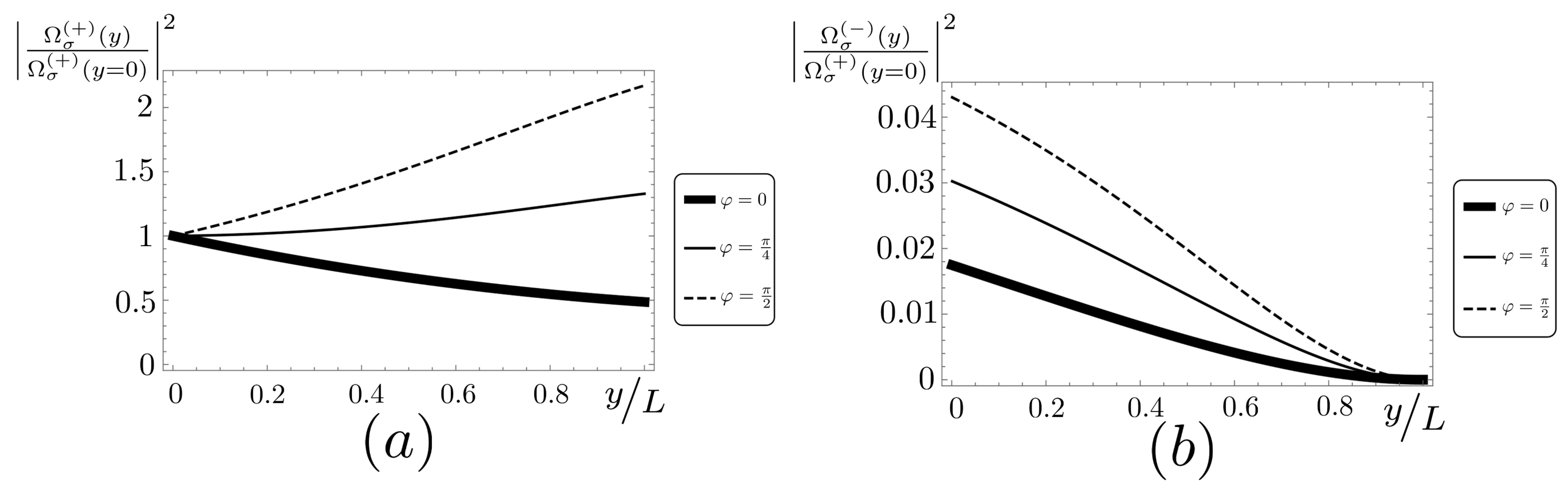}
\par\end{centering}
\caption{Probe field intensities (a) $\left|\frac{\Omega_{\sigma}^{+}\left(y\right)}{\Omega_{\sigma}^{+}\left(0\right)}\right|^{2}$
and (b) $\left|\frac{\Omega_{\sigma}^{-}\left(y\right)}{\Omega_{\sigma}^{+}\left(0\right)}\right|^{2}$
as functions of the (normalized) propagation distance $\frac{y}{L}$
for $\varphi=0$ (full thick line), $\frac{\pi}{4}$  (full line), $\frac{\pi}{2}$ (dashed line).
\label{Fig8}}
\end{figure}

\begin{figure}
\begin{centering}
\includegraphics[width=\textwidth]{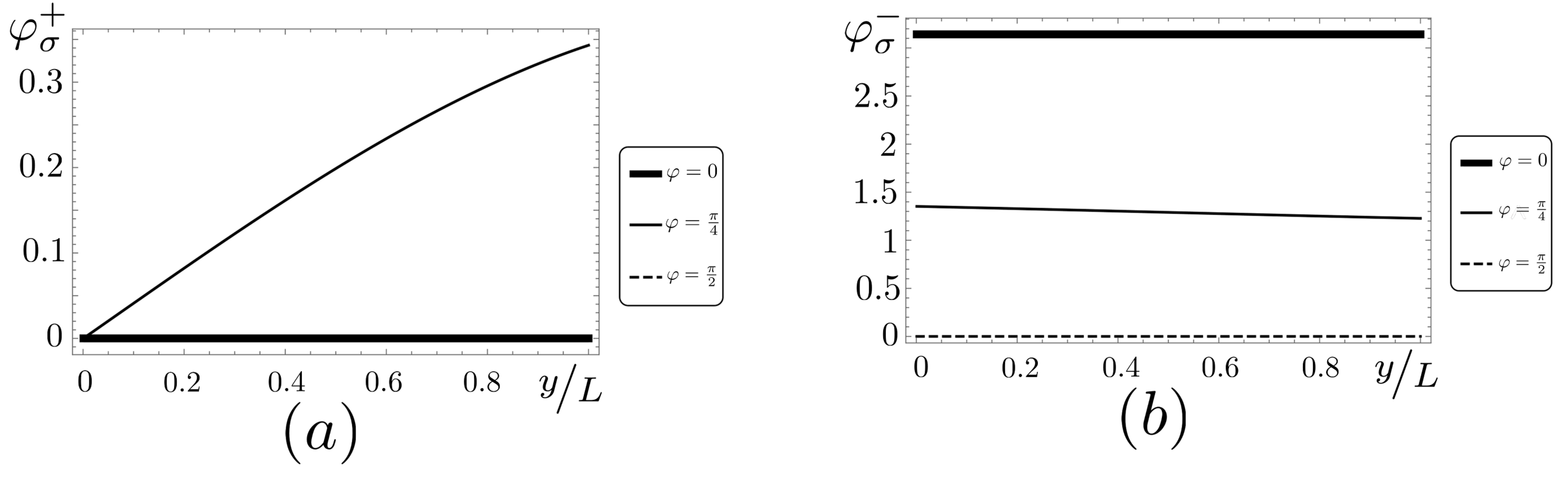}
\par\end{centering}
\caption{Accumulated phases (a) $\varphi_{\sigma}^{+}\left(y\right)$ of the
forward-propagating probe field and (b) $\varphi_{\sigma}^{-}\left(y\right)$
of the backward-propagating probe field as functions of the (normalized)
propagation distance $\frac{y}{L}$ for $\varphi=0$  (full thick line), $\frac{\pi}{4}$  (full line), $\frac{\pi}{2}$ (dashed line).\label{Fig9}}
\end{figure}

In Fig. \ref{Fig8} are plotted the intensities of (a) the forward
probe field, $\left|\frac{\Omega_{\sigma}^{+}\left(y\right)}{\Omega_{\sigma}^{+}\left(0\right)}\right|^{2}$,
and (b) the backward probe field, $\left|\frac{\Omega_{\sigma}^{-}\left(y\right)}{\Omega_{\sigma}^{+}\left(0\right)}\right|^{2}$,
as functions of the normalized coordinate $\frac{y}{L}$ for three
values of the relative phase $\varphi=0,\frac{\pi}{4},\frac{\pi}{2}$,
and $\Delta_{0}=0$, $\frac{\Omega_{\pi}^{+}\left(0\right)}{\Gamma_{d}}=0.4$,
$\frac{\Omega_{\pi}^{-}\left(L\right)}{\Gamma_{d}}=0.16$, $\alpha_{0}L=0.6$.
In Figs. \ref{Fig9} (a,b) %are 
 we  plotted the corresponding accumulated
phases $\varphi_{\sigma}^{\pm}\left(y\right)$. As shown  in  %by 
Fig. \ref{Fig8}-a, $\varphi=0$ leads to absorption of the forward component of the probe
whereas $\varphi=\frac{\pi}{2}$ leads to its significant amplification.
The case $\varphi=\frac{\pi}{4}$ corresponds to a small modification
of the probe. This is in  agreement  %line 
with the results of the numerical simulations
(Fig.\ref{Fig5}-a). We also %notice 
 note  that  the  intensity depends quasi-linearly
on the propagation distance which  confirms  %comforts 
the linear approximation %performed 
 made  in Eqs. (\ref{Omegasigma+final},\ref{TransmissionApprox}).
In Fig. \ref{Fig8}-b, %is represented  
the backward probe component  is shown 
which vanishes at $y=L$. The field is amplified from the  output  %exit 
to
the  input %entrance 
of the sample. Its intensity at the  input %entrance 
depends %significantly 
  strongly  on the dephasing. On the other hand, the shape of the
curves %is almost parabolic 
around $y=L$    is almost parabolic. In fact, %. Indeed, 
when $y$ is close
to $L$, Eq. (\ref{Omegasigma-final}) yields $\left|\Omega_{\sigma}^{-}\left(y\right)\right|^{2}\approx\left|\Omega_{\sigma}^{+}\left(0\right)\right|^{2}\left|c_{+-}\right|^{2}\alpha_{0}^{2}\left(y-L\right)^{2}$. 

The accumulated phases $\varphi_{\sigma}^{+}\left(y\right)$ and $\varphi_{\sigma}^{-}\left(y\right)$
are   shown %represented 
in Fig.\ref{Fig9}-a and b, respectively. The forward
component exhibits a vanishing phase at the   input %entrance 
and accumulates
a significant value along   the  propagation $\left(\varphi_{\sigma}^{+}\left(y=L\right)\approx0.34\right)$
for $\varphi=\frac{\pi}{4}$ whereas for $\varphi=0$ and $\varphi=\frac{\pi}{2}$
the phase remains equal to zero. This effect can be explained   by %from
the simple analytical model introduced in Subsec. \ref{SmallOD} which
leads to Eq. (\ref{Omegasigma+final}). We then have $\varphi_{\sigma}^{+}\left(y\right)\approx-\alpha_{0}y\left|c_{+}\right|\cos\left(2\varphi+\varphi_{L}\right)$.
At resonance, $\varphi_{L}=\frac{\pi}{2}$ and   therefore %thus 
 $\varphi_{\sigma}^{+}\left(y\right)=0$
for $\varphi=0,\frac{\pi}{2}$ and $\varphi_{\sigma}^{+}\left(y\right)\approx\alpha_{0}y\left|c_{+}\right|$
for $\varphi=\frac{\pi}{4}$.   With %Using 
$\left|c_{+}\left(0\right)\right|=0.683$,
the analytical model predicts an accumulated phase $\varphi_{\sigma}^{+}\left(y=L\right)\approx0.41$
to compare with the value $0.34$  %given by 
  obtained from  exact numerical simulations.
The backward component %has 
  shows  a different   behaviour. In fact, %behavior. Indeed, 
 due to the
boundary condition $\Omega_{\sigma}^{-}\left(L\right)=0$, the value
of the phase $\varphi_{\sigma}^{-}$ cannot be set either at the   input %entrance
$\left(y=0\right)$ or  at  the  output  %exit 
$\left(y=L\right)$ of the sample,
only the propagation dynamics allows us to recover it. From Eq. (\ref{Omegasigma-final}),
$\varphi_{\sigma}^{-}\left(y\right)\approx\pi-2\varphi+\alpha_{0}\left(y-L\right)\left|c_{+}\right|\cos\left(2\varphi+\varphi_{L}\right)$,
which   yields %leads 
$\varphi_{\sigma}^{-}\left(L\right)\approx\pi,\frac{\pi}{2},0$
for $\varphi=0,\frac{\pi}{4}$ and $\frac{\pi}{2}$ in   accordance %agreement
with the numerical values read on inset (b) of Fig. 9, \emph{i.e.}
$3.14,1.35,0$, respectively.

\section{Conclusion}

In this article, the  dynamical behaviour  of an ensemble of duplicated two-level
systems driven by a stationary control field and subject to
a weak probe field was  studied in detail.
In particular, the expressions
of the probe reflection, $R$, and transmission, $T$, factors were
determined and their dependence on   the field parameters was %investigated
 studied.
We suggested an experimentally feasible implementation of our scheme involving transitions in  $^6$Li which can be excited with conventional laser beams.
The values for the optical depth and Rabi frequencies assumed in our simulations can be easily achieved experimentally. Indeed, 
since  $\Gamma \approx 0.45$MHz,  $\alpha_{0}L = 1$ can be obtained  for $L=1$cm and an atomic density $N=3\times10^8 \text{at}\cdot \text{cm}^{-3}$. Rabi frequency  $\Omega_\pi=\Gamma$ can also be reached for control field intensities of the order of  $0.2\text{mW}\cdot\text{cm}^{-2}$.

Both amplification and absorption were shown to be achievable. The control potentialities offered by this system were investigated, in particular the %situation of
parametric regime obtained for moderate optical depths
which leads to proper phase control over $R$ and $T$. For arbitrarily higher optical depths,
both the driving fields and the phase accumulated by the probe field
along its propagation are affected %which 
drastically %modified 
 modifying  the
behavior of the system. Previous works on co-propagating control and
probe field schemes have already shown new interesting phenomena
%like 
such as  phase saturation and transparency \cite{HB08b}. This paves
the way  for  %to 
further investigations for the present stationary control
field configuration. Numerical simulations also showed a strong increase  in %of the 
reflection and transmission factors with %the 
optical depth and  also  a \emph{phase-sensitivity}.  % also.
The combination of these effects %offers 
 provides  a stimulating motivation for
further study of this case. In particular, we believe that (i) higher reflectivity will be achievable and that (ii) it may be possible to design the pump standing-wave  so that the %in such a way that 
reflection and transmission factors reach their maxima for different values of the relative phase – ideally, transmission would be maximal when reflection is minimal, and vice versa, thus implementing a phase-sensitive optical switch.

\section*{Acknowledgements}
This research was funded in part by l'Agence Nationale de la Recherche (ANR), Project ANR-22-CE47-0011. For the purpose of open access, the authors have applied a CC-BY public copyright licence to any Author Accepted Manuscript (AAM) version arising from this submission.



%% For example:

%% \bibitem[Aladro et al.(2015)]{Aladro15} Aladro, R., Martín, S., Riquelme, D., et al. 2015, \aas, 579, A101

%%\end{thebibliography}

\end{document}